# Spin-orbit magnetic state readout in scaled ferromagnetic/heavy metal nanostructures

Van Tuong Pham[1,†], Inge Groen[1,†], Sasikanth Manipatruni[2], Won Young Choi[1], Dmitri E. Nikonov[2], Edurne Sagasta[1], Chia-Ching Lin[2], Tanay Gosavi[2], Alain Marty[3], Luis E. Hueso[1,4], Ian Young[2], and Fèlix Casanova [1,4,*].

[1] CIC nanoGUNE, 20018 Donostia-San Sebastián, Basque Country, Spain.

[2] Components Research, Intel Corp., Hillsboro, OR 97124, USA.

[3] SPINTEC, CEA-INAC/CNRS/Univ. Grenoble Alpes, F-38000, Grenoble, France.

[4] IKERBASQUE, Basque Foundation for Science, 48013 Bilbao, Basque Country, Spain.

*Correspondence to: f.casanova@nanogune.eu.

† These authors contributed equally to this work.

**Abstract:** Efficient detection of the magnetic state at nanoscale dimensions is an important step to utilize spin logic devices for computing. Magnetoresistance effects have been hitherto used in magnetic state detection, but they suffer from energetically unfavorable scaling and do not generate an electromotive force that can be used to drive a circuit element for logic device applications. Here, we experimentally show that a favorable miniaturization law is possible via the use of spin-Hall detection of the in-plane magnetic state of a magnet. This scaling law allows us to obtain a giant signal by spin Hall effect in CoFe/Pt nanostructures and quantify an effective spin-to-charge conversion rate for the CoFe/Pt system. The spin-to-charge conversion can be described as a current source with an internal resistance, *i.e.*, it generates an electromotive force that can be used to drive computing circuits. We predict that the spin-orbit detection of magnetic states can reach high efficiency at reduced dimensions, paving the way for scalable spin-orbit logic devices and memories.

Modern computing transistor technology is scaled to tens of nanometers[1] in lateral dimensions driven by the favorable miniaturization (Moore's Law)[2]. Such a favorable miniaturization[3] is an essential requirement for enabling spin logic[4-7] in computing but it has so far been a missing focus in spintronics. In particular, energy efficient detection of the magnetic state at the nanoscale dimensions is an important step to realize spin logic devices for computing. Up to now, magnetic state sensing techniques have relied on magnetoresistances such as anisotropic magnetoresistance (AMR)[8], giant magnetoresistance (GMR)[9,10], colossal magnetoresistance (CMR)[11], and tunneling magnetoresistance (TMR)[12]. Even if TMR has been steadily improved to large values (>1000%)[13], the magnetoresistance techniques are unfavorable in terms of energy for sensing a magnetic state because the resistance of the device increases quadratically when scaling down the area of the device[14]. Also, importantly, magnetoresistance techniques cannot generate an electromotive force (i.e., an electric current) that can be used to drive another circuit element, a requirement for a spintronic logic device realization[15] severely limiting their potential for computational logic. Hence, it is of great interest to develop a magnetic state detection mechanism that scales favorably with dimensions and is able to drive current for logic in computing applications.

The discovery of new spin-orbit coupling (SOC) phenomena leading to spin-to-charge current conversion (SCC), such as the spin Hall effect (SHE) in bulk materials[16, 17] and the Edelstein effect in Rashba interfaces[18] and topological insulators[19, 20] is expanding the potential for a second generation of spintronic devices that can integrate high-density memory with high speed operations[21]. In particular, spin-orbit torque experiments driven by the SHE and/or the Edelstein effect in heavy metals or their interfaces can be used to write magnetic memories[22, 23]. Since a high spin current density is required to generate efficient torques, most efforts in the studies of SHE have focused on conducting materials with high SCC efficiency[24]. Notably, not only the writing but also the reading process is feasible by using SCC. The reciprocal phenomena [inverse SHE (ISHE), inverse Edelstein effect] can potentially be used to read the magnetic state of a magnetic element, as recently proposed for a magnetoelectric spin-orbit (MESO) based logic[4,15]. In fact, both the ISHE and SHE can detect the magnetization state of a ferromagnetic (FM) element in a nanostructure: the ISHE probes the spin current electrically injected from the FM element by converting it into a charge current; reciprocally, the SHE generates the spin current that is probed by the spin-dependent electrochemical potential[25] of the FM element. Typical examples of metallic nanostructures that use the SHE/ISHE for spin injection/detection are the lateral spin valves developed to quantify the SHE via nonlocal techniques[26]. In these devices, the FM electrode and the SHE electrode are separated by a non-magnetic (NM) channel with weak SOC that transports the spin currents between them. However, the spin Hall signals (~0.1−1 mΩ) are small due to the spin current diffusion along the NM channel, spin losses at the two interfaces, and the shunting effect of the NM channel on the SHE electrode[26]. A prerequisite to read the magnetic state of an FM element in potential applications such as the MESO logic is a sufficiently large spin Hall signal (~10 kΩ)[4,15]. Local spin injection/detection techniques in simpler FM/NM nanostructures could help bridging this large gap, but reported values are still far away (~1-10 mΩ)[25,27].

Here, we demonstrate the favorable miniaturization of the spin-orbit-based readout of a magnetic state. We show that the spin-to-charge conversion using ISHE allows us to independently enhance the output voltage (needed to read the in-plane magnetization) and the output current (needed for cascading circuit elements) with downscaling of different device dimensions, which are necessary conditions for implementing spin-based logic such as the MESO logic[4]. We exploit these unveiled scaling laws to obtain giant spin Hall signals at room temperature (0.3 Ω) in simple local CoFe/Pt nanostructures, which arise from the small dimensions and high resistivities of Pt and CoFe, whereas the effective spin-to-charge conversion rate ($\lambda_{eff}$) remains constant for the CoFe/Pt system. We show that it is possible to further increase the spin Hall signal to the values needed to operate the MESO logic with the proper choice of materials, making it a real candidate for post-CMOS computing.

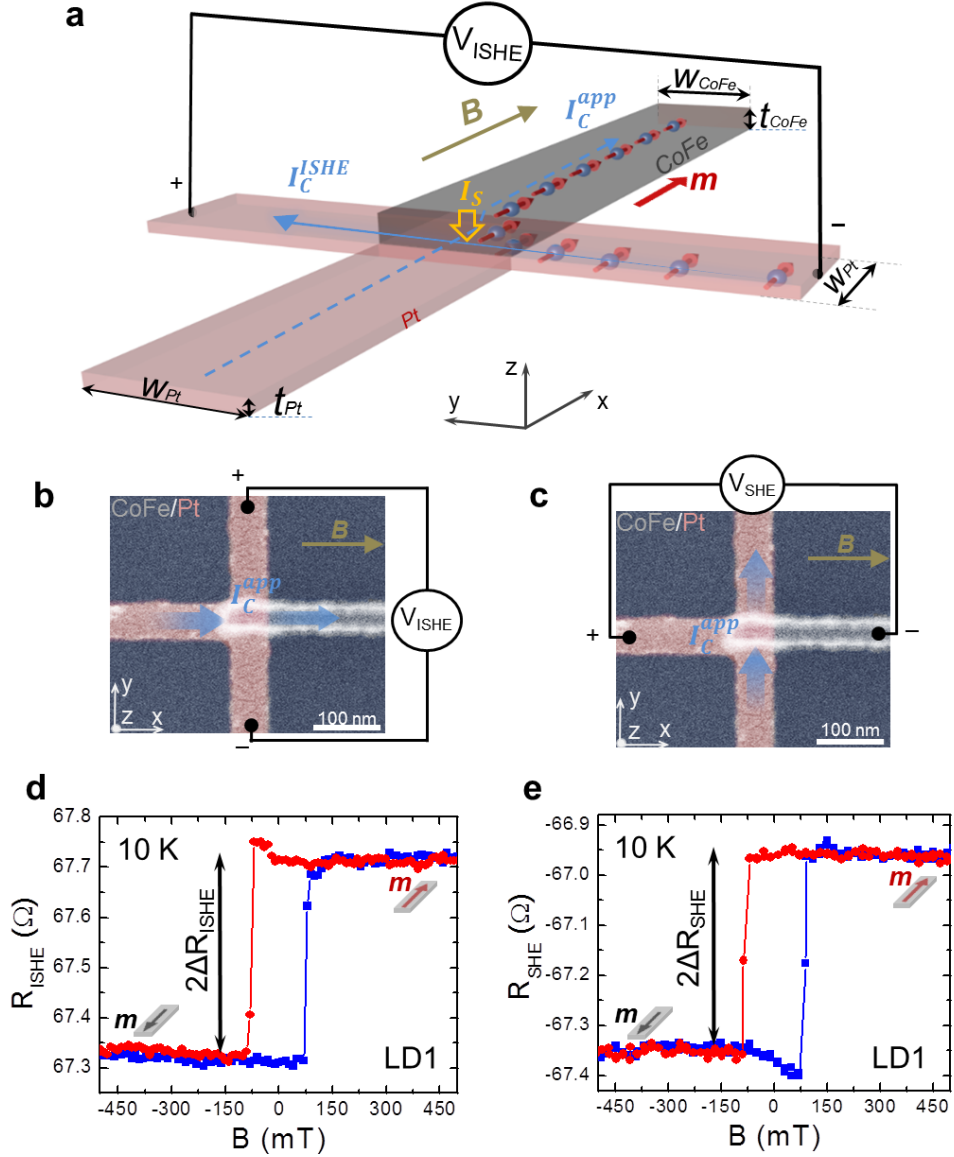

**Fig. 1. Sketch, images and measurements of the spin-to-charge conversion device used for in-plane magnetic state detection. a**, CoFe electrode (grey) is patterned on top of a Pt T-shaped nanostructure (red). The magnetization (***m***) of the CoFe electrode (red arrow) is controlled by a magnetic field ***B*** (gold arrow). A charge current $I_C^{app}$ (dashed blue arrow) is used to inject a spin-polarized current $I_S$ along the $z-$direction (yellow arrow) from CoFe into Pt. $I_S$ is converted into a charge current $I_C^{ISHE}$ (solid blue arrow) in $y-$direction by the ISHE in Pt, which is measured as a voltage ($V_{ISHE}$) in the open circuit condition. The sign of the voltage is changed by ***m*** reversal. Note that the spin-polarized current $I_S$ is simplified as a subtraction between flows of the majority and the minority spins. **b** and **c** show an SEM image of device LD1 with the measurement configurations for the ISHE (spin-to-charge conversion) and the SHE (charge-to-spin conversion), respectively. **d** and **e** are the transverse resistances $R_{ISHE}$ and $R_{SHE}$, respectively, as a function of the magnetic field measured for device LD1 at 10 K with the ISHE and SHE configurations as shown in **b** and **c**, respectively. The two magnetization orientations are indicated. Blue squares correspond to trace of the magnetic field and red circles to retrace.

The fabrication process and measurement setup of the devices is described in Methods. Figure 1a depicts the design of the device, simply comprising a single CoFe electrode connected on top of a Pt T-shaped nanostructure, with the sketch for the ISHE measurement. A scanning electron microscopy (SEM) image of one of the devices (LD1) is shown in Figs. 1b and 1c, where the measurement configurations for the ISHE and for the SHE, respectively, are also presented. In the ISHE measurement, the transverse charge current is generated from the spin-polarized current injected from CoFe into Pt by applying a bias ($I_C^{app}$) through the CoFe/Pt interface. The conversion from spin current density ($J_S$) to charge current density ($J_C^{ISHE}$) in the Pt can be expressed as $J_C^{ISHE} = \left[\frac{e}{\hbar}\right]\theta_{SH} J_S \times s$, where $s$ is the spin-polarization vector of the spin current, which is determined by the magnetization vector ($m$) of the CoFe. $\theta_{SH}$ is the spin Hall angle of Pt. The produced charge current is detected as a voltage ($V_{ISHE}$) in the open circuit condition, which will change sign upon reversal of $m$. Reciprocally, in the SHE measurement, we apply a charge current ($I_C^{app}$) along the transverse wire of the Pt T-shaped nanostructure (*y*-direction) to generate a pure spin current along *z*-direction polarized in the *x*-direction, which leads to a spin accumulation at the CoFe/Pt interface. This spin accumulation is then probed by the magnetization of the CoFe electrode. The Fermi level of the CoFe electrode aligns with the electrochemical potential of the majority (minority) spins of the spin accumulation when its magnetization is along +*x* (–*x*), creating a net positive (negative) interface voltage ($V_{SHE}$). Consequently, in-plane magnetization reversal will lead to a sign change of the voltage.

The main results for device LD1 ($t_{Pt}$ = 6 nm, $t_{CoFe}$ = 10 nm and $w_{Pt}$ = 60 nm) are shown in Figs. 1d-1e. We plot the transverse resistances as a function of the magnetic field at 10 K for the ISHE (Fig. 1d) and the SHE (Fig. 1e) configurations, which are obtained from the ratio between the measured voltages and the applied current ($R_{(I)SHE} = V_{(I)SHE}/I_C^{app}$). We define the spin Hall signal as the difference in the resistance between the two magnetization orientations ($2\Delta R_{(I)SHE}$). The magnetic hysteresis loops of $R_{ISHE}$ and $R_{SHE}$ are reciprocal, following the Onsager relations. The two magnetization configurations of the electrode are clearly distinguished by the sharp resistance jump at the switching fields, as expected from the mechanism described above. The switching fields are similar to previous reports on 50-nm-wide CoFe wires[25]. This giant signal is reproducible, as shown in another device with similar geometry (Supplementary Note 1).

The hysteresis loops stay well defined up to room temperature (see Fig. 2a). Most importantly, the obtained spin Hall signals [300 mΩ (410 mΩ) at 300 K (10 K)] are three orders of magnitude higher than those measured in nonlocal measurements using Py/Cu/Pt lateral spin valves [0.3 mΩ (0.15 mΩ) at 10 K (300 K)] [26]. An enhancement of the spin Hall signal is expected in the local spin detection technique, because it avoids the spin diffusion as well as the shunting effect in the Cu channel. Although these issues can be solved by replacing Cu with graphene[28], the simpler design of the local spin detection allows for the downscaling of the metallic nanostructure to very low dimensions (device LD1 has a 6-nm-thick and 60-nm-wide Pt wire) and reproducibility for large scale fabrication. Remarkably, the spin Hall signals are ~20 times larger than those measured locally in related CoFe/Pt nanostructures (12 mΩ at 300 K)[25] and two orders of magnitude larger than those observed in a larger CoFeB/MgO/Pt structure (~1 mΩ at 300 K)[27], because we take advantage of the lower dimensions and higher resistivities of our nanostructure. In order to directly confirm this advantage, the hysteresis loop of $R_{ISHE}$ is measured in a larger device (LD2, with $t_{Pt} = 10$ nm, $t_{CoFe} = 19$ nm and $w_{Pt} = 160$ nm) with lower resistivities, which is plotted in Fig. 2b. The spin Hall signal (14 mΩ at 300 K), as expected, is much smaller than that of device LD1. The temperature dependence of the spin Hall signals $2\Delta R_{(I)SHE}$ for devices LD1 and LD2 are plotted in Figs. 2c and 2d, respectively.

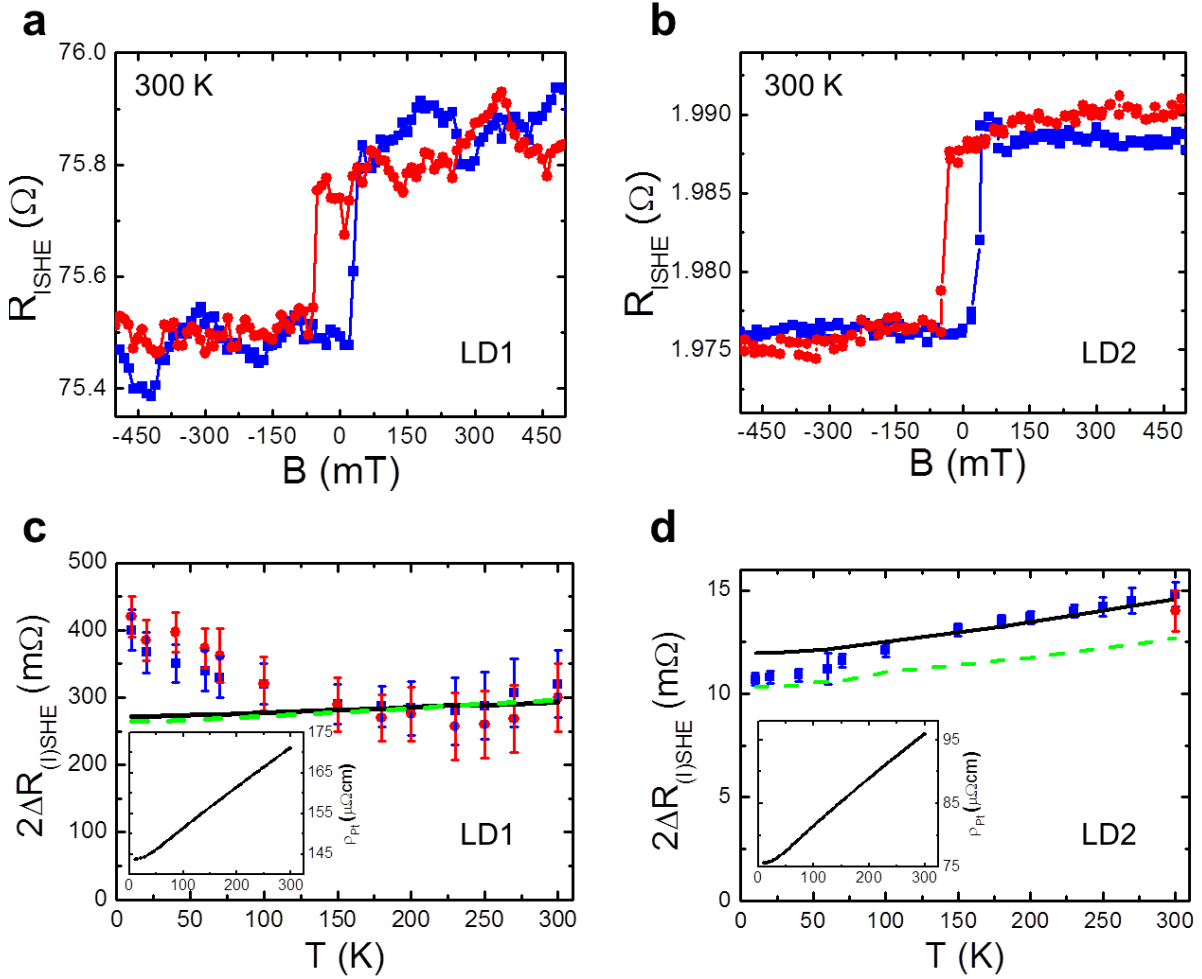

**Fig. 2. Temperature dependence of the spin Hall signals. a** and **b** are the ISHE transverse resistances ($R_{ISHE}$) as a function of the magnetic field (trace and retrace) measured at room temperature for the devices LD1 and LD2, respectively. **c** and **d** plot the direct ($2\Delta R_{SHE}$, red circles) and inverse ($2\Delta R_{ISHE}$, blue circles) spin Hall signals as a function of the temperature for the devices LD1 and LD2, respectively. Error bars are calculated using the s.d. associated with the statistical average of $R_{ISHE}$ in both the positive and negative **m** states. The black solid line is a calculation of $2\Delta R_{(I)SHE}$ vs temperature based on the 1D spin diffusion model [equation (2)]. The green dashed line is a 3D finite-element method simulation of $2\Delta R_{(I)SHE}$ vs temperature based on the spin diffusion model, which also include the contribution of the anomalous Hall effect in the CoFe. Insets show the measured Pt resistivities for each device.

In order to quantitatively explain the obtained spin Hall signals, we apply the one-dimensional (1D) spin diffusion model. Following Ref. 25, the spin Hall signals arising from the device geometry of Fig. 1a can be written as a simple expression:

$$\Delta R_{(I)SHE} = \frac{P_{CoFe}\theta_{SH}\lambda_{Pt}}{\left(\frac{t_{CoFe}}{\rho_{CoFe}}+\frac{t_{Pt}}{\rho_{Pt}}\right)w_{Pt}} \times \frac{1-\frac{1}{\cosh(t_{Pt}/\lambda_{Pt})}}{\tanh(t_{Pt}/\lambda_{Pt})+\frac{\lambda_{Pt}\rho_{Pt}}{\lambda_{CoFe}\rho^*_{CoFe}}\tanh(t_{CoFe}/\lambda_{CoFe})} \quad (1)$$

with $\rho^*_{CoFe} = \rho_{CoFe}/(1-P^2_{CoFe})$, where $P_{CoFe}$, $\lambda_{Pt,CoFe}$ and $\rho_{Pt,CoFe}$ are the spin polarization of CoFe, the spin diffusion length and the resistivity, respectively. The subscripts denote the materials. The geometrical parameters ($t_{Pt,CoFe}$, $w_{Pt}$) are presented in Fig. 1a. For details of the 1D model, see Supplementary Note 2. For the calculation of the spin Hall signal, the material parameters based on the measured resistivities in the devices, which are plotted at the insets of Figs. 2c and 2d, are considered. In particular, the value of spin Hall angle of Pt, which has an intrinsic and extrinsic (skew scattering) contribution, is given by $\theta_{SH} = \sigma_{SH}^{int}\rho_{Pt} + \alpha_{SS}$, where $\sigma_{SH}^{int} = 1600$ [ℏ/e] $\Omega^{-1}cm^{-1}$ is the intrinsic spin Hall conductivity and $\alpha_{SS} = 0.02$ is the skew scattering angle[26]. The value of the spin diffusion length of Pt is also determined by its resistivity ($\rho_{Pt}\lambda_{Pt} = 0.77$ f$\Omega m^2$)[29], because the spin relaxation in Pt is dominated by the Elliott-Yafet mechanism[26,29]. Similarly, the spin diffusion length of CoFe is calculated by using $\rho_{CoFe}\lambda_{CoFe} = 1.29$ f$\Omega m^2$ [30]. Finally, we take $P_{CoFe} = 0.48$ at both 10 K and 300 K from G. Zahnd et al.[30].

The calculated spin Hall signal as a function of temperature using the 1D model [equation (1)] is plotted as a black solid line in Figs. 2c (device LD1) and 2d (device LD2). It is worth noting that this is not a fit to the experimental data, but a calculation from independent parameters. The agreement between the calculation and the experimental data, especially good above 150 K, confirms the SHE of Pt as the origin of the signal. The deviation at lower temperatures could be caused by slight mismatches between the used parameters of the literature and the actual parameters. Alternatively, the presence of a strong interfacial spin-to-charge conversion at the FM/Pt interface[31] could be a relevant contribution in the thinnest Pt, an effect that might be more significant at low temperatures[32]. For a more accurate calculation, we use a 3D finite-element-method (FEM) to calculate the spin Hall signal, plotted as a green dashed line in Figs. 2c and 2d (Supplementary Note 3). Although the 3D FEM accounts for the current distribution in the nanostructures, whereas the 1D model does not, it is worth noting that the calculated spin Hall signals for device LD1 (Fig. 2c) are almost identical, because the two models are equivalent in this thin and narrow nanostructure. However, the difference between the two models becomes more apparent in the larger nanostructure of device LD2, where the signals from the 3D model are lower than the ones from the 1D model (Fig. 2d).

In order to rule out possible artifacts that could mimic the spin Hall signals, a careful analysis is performed. The most probable spurious effect is the anomalous Hall effect (AHE)[31] which could originate in the CoFe by the geometry of the applied current lines, such that an anomalous Hall signal would appear in the same shape simultaneously with the spin Hall signal. For this reason, we measured the anomalous Hall angle ($\theta_{AH}$) in a cross fabricated in the very same device. The measured $\theta_{AH}$ shows rather small values (0.44–0.57%), in line with previous results[25] (Supplementary Note 4). Since the 1D model cannot account for an anomalous Hall contribution in the overall signal due to symmetry ($I_C^{app}$ and $\boldsymbol{m}$ in the FM are along the same direction), the 3D FEM model is used to include the contribution of the AHE implementing the experimentally obtained $\theta_{AH}$. In this case, a finite anomalous Hall contribution is allowed when the charge current distribution at the CoFe/Pt interfacial region is considered. The anomalous Hall contribution, which is already included in the calculated signal (green dashed line in Figs. 2c and 2d) is less than 5% of the total signal in both devices. Other spurious effects such as AMR, thermal-related effects such as spin Nernst and spin Seebeck effects[34], and ordinary Hall effect due to the stray field of the FM electrode can be neglected as shown in Ref. 25. Other spurious signals such as planar Hall effect (Supplementary Note 5) and spin caloritronic effects[34] (Supplementary Note 6) are carefully considered and shown to be negligible contributions to the spin Hall signal. Furthermore, a control experiment in a CoFe/Ta sample shows a negative spin Hall signal (Supplementary Note 7), expected from the negative spin Hall conductivity in Ta,[35] confirming that all the results presented in this work are primarily due to the spin Hall effect.

Extended Data Table 1 lists the experimental spin Hall signals measured at room temperature in devices with different geometries. The devices are fabricated on seven different $SiO_2$/Si substrates (LD1-LD7). The resistivities and all geometrical parameters are obtained from the very same device where the spin Hall signal is measured. The set of devices with varying $w_{Pt}$ (LD3) and $w_{CoFe}$ (LD4) allows us to confirm the dependence of the spin Hall signal with these two relevant dimensions of the device given by equation (1): whereas $2\Delta R_{(I)SHE} \propto w_{Pt}^{-1}$, it does not depend on $w_{CoFe}$ (Supplementary Note 8).

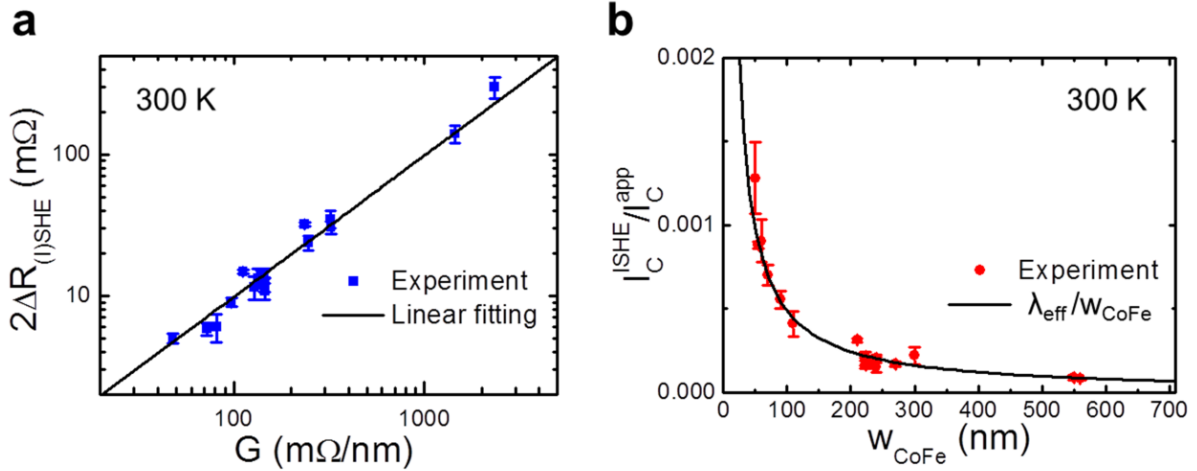

**Fig. 3. Favorable scaling law for the spin-to-charge conversion rates. a**, Measured spin Hall signals at room temperature for the different devices shown in Extended Data Table 1 plotted as a function of the "geometrical" factor $G$ (blue solid squares). The black solid line is a linear fit to Eq. 2, with a slope that corresponds to two times of the "efficiency" factor $\lambda_{eff}$ (in nanometers). **b**, The net charge current conversion rate, extracted from the data in **a**, as a function of the width of the CoFe electrode (red solid circles). The black solid line is an independent plot of equation (3) (not a fit) obtained by fixing $\lambda_{eff}$ to the value previously extracted from the linear fit in **a**. Error bars are calculated using the s.d. associated with the statistical average of $R_{ISHE}$ in both the positive and negative $\boldsymbol{m}$ states.

Next, we will use the 1D model to perform a comprehensive analysis on the data of all the devices. For systems composed of materials with much shorter spin diffusion lengths than their thicknesses, which is the case in this work ($\lambda_{Pt} \ll t_{Pt}$, $\lambda_{CoFe} \ll t_{CoFe}$), a simplified expression for equation (1) is given by:

$$\Delta R_{(I)SHE} = G \times \lambda_{eff} \qquad (2)$$

where $G = \frac{1}{\left(\frac{t_{CoFe}}{\rho_{CoFe}} + \frac{t_{Pt}}{\rho_{Pt}}\right)w_{Pt}}$ and $\lambda_{eff} = \frac{P_{CoFe}\theta_{SH}\lambda_{Pt}}{1 + \frac{\lambda_{Pt}\rho_{Pt}}{\lambda_{CoFe}\rho^*_{CoFe}}}$ (Supplementary Note 2). The "geometrical" factor $G$ is related to the transverse resistance of the device given by the dimensions and the resistivities of the CoFe and Pt, which can be measured directly from the devices. The "efficiency" factor $\lambda_{eff}$ is given by the material properties of the system: the SCC efficiency in Pt ($\theta_{SH}\lambda_{Pt}$ product, which is essentially constant[26]), the spin polarization of CoFe ($P_{CoFe}$), and the spin resistance mismatch of the two materials ($\lambda_{Pt}\rho_{Pt}/\lambda_{CoFe}\rho^*_{CoFe}$). Hence, $\lambda_{eff}$ is expected to be constant for a specific metallic FM/NM system for any NM and FM resistivity. The scaling law presented in equation (2) can be understood by considering an equivalent circuit which is adapted from the transmission model for ISHE introduced by Sayed et al.[36] The factor $G$ would correspond to the transverse resistance per unit length of the internal resistance and $\lambda_{eff}$ would be the effective length along which the ISHE current is generated (Supplementary Note 9 for a detailed discussion). Experimentally, a linear relation between the spin Hall signals and $G$ is shown in Fig. 3a, confirming the prediction of a constant $\lambda_{eff}$ for the overall set of CoFe/Pt devices. A linear fit of the data to equation (2) (black solid line) yields $\lambda_{eff}$ = 0.05 ± 0.01 nm. This scaling plot confirms that the obtained giant spin Hall signal in device LD1 is given by the large value of $G$ (2.34 Ωnm$^{-1}$), achieved due to its small dimensions and high resistivities. This is thus a guideline that can be followed to obtain even higher spin Hall signals. Indeed, we achieve a spin Hall signal of 3.4 Ω by adapting the device technique to a new spin-orbit system, CoFe/Ta, with larger values of $G$ (17.8 Ωnm$^{-1}$) and $\lambda_{eff}$ (0.10 nm), as shown in Supplementary Note 7.

It is worth mentioning that the excellent match between the experimental values and the model given by equation (2) makes this local spin detection technique a promising alternative to quantify SCC efficiency, with a data analysis much simpler than other widely used local techniques such as harmonic Hall voltage measurements[37], spin torque-ferromagnetic resonance[23, 29], or spin pumping[20].

The spin Hall signal $\Delta R_{(I)SHE}$ quantifies the rate of the transverse voltage output to the charge current input ($I_C^{app}$). The giant values obtained in our optimized local devices allow us to read the in-plane magnetization state of a FM electrode, an essential ingredient for the MESO logic. However, the performance of the MESO logic also relies on the output current, as it is used to charge/discharge a magnetoelectric capacitor node to enable switching of the next logic gate[4, 15]. Particularly, whereas the produced voltage directly determines the capability of switching the next FM element in the circuit with the magnetoelectric effect in the MESO concept[4], the produced current defines the switching energy and delay time. In this regard, the rate of the transverse charge current output ($I_C^{ISHE}$) to $I_C^{app}$ should also be maximized to minimize the switching energy. This rate can be expressed as (Supplementary Note 2):

$$\frac{I_C^{ISHE}}{I_C^{app}} = \frac{1}{w_{CoFe}} \times \lambda_{eff} . \qquad (3)$$

In this case, the only difference with $\Delta R_{(I)SHE}$ is the "geometrical" factor, which is simply given by $\frac{1}{w_{CoFe}}$ instead of $\frac{1}{\left(\frac{t_{CoFe}}{\rho_{CoFe}} + \frac{t_{Pt}}{\rho_{Pt}}\right)w_{Pt}}$. Importantly, this shows that, in a specific material system, the ISHE outputs in current and in voltage can be independently optimized by tuning different dimensions, *i.e.*, the ISHE current source and its internal resistance controlled by different scaling laws (Supplementary Note 9). The spin Hall signals for the different devices are converted into $I_C^{ISHE}/I_C^{app}$ and plotted as a function of $w_{CoFe}$ in Fig. 3b. A clear enhancement of the charge current conversion rate by downscaling the width of CoFe wire is shown. The largest value $I_C^{ISHE}/I_C^{app} = 1.3 \times 10^{-3}$ is achieved in the smallest $w_{CoFe}$= 50 nm. In the same figure, we can directly plot equation (3) with no fitting parameters by simply using the previously obtained $\lambda_{eff}$ for our system, that shows a perfect match with the experimental data. It is worth to emphasize that the downscaling of the FM width, in addition to the enhancement of the output current, also favors the reduction of the switching energy in the MESO device [4].

The results in Fig. 3b indicate that the requirements for improving the output charge current for MESO-based logic devices are (i) downscaling the FM nanostructure ($w_{CoFe}$) and (ii) using a FM/NM system with a large $\lambda_{eff}$. Considering that the spin polarization of any magnetic material $P_{FM}$ cannot be larger than 1, $\lambda_{eff}$ can be mainly enhanced by the SCC efficiency in the NM, *i.e.,* the $\theta_{SH}\lambda_{NM}$ product in bulk materials with strong SOC or the inverse Edelstein length $\lambda_{IEE}$ at Rashba interfaces or in topological insulators. Interestingly, the SCC efficiency can reach several nanometers in systems such as $Bi_2Se_3$ ($\theta_{SH}\lambda_{NM}$ ~ 10 nm),[20] graphene/$MoS_2$ ($\theta_{SH}\lambda_{NM}$ ~ 10 nm),[38] α-Sn ($\lambda_{IEE}$ ~ 2.1 nm),[21] or $LaAlO_3$/$SrTiO_3$ ($\lambda_{IEE}$ up to 6.4 nm, tunable with gate voltage)[39]. Therefore, based on equation (3), by reducing the CoFe electrode width to few tens of nm and replacing Pt with one of these novel systems with large SCC efficiency, the $I_C^{ISHE}/I_C^{app}$ ratio could potentially approach one.

In conclusion, this work is the first experimental step towards the realization of the MESO logic proposal[4]. The observation of giant spin Hall signals (300 mΩ at 300 K) in the CoFe/Pt metallic system with a local spin detection scheme demonstrates the feasibility of the spin-orbit-based reading in the MESO logic devices[4,15]. The understanding of the scaling laws that shows favorable miniaturization and the role of the material parameters for optimal spin-to-charge current conversion provide guidelines for optimizing the spin Hall signals by choosing the proper geometry and the best material systems for future devices. We anticipate that spin-momentum locking in topological insulators[19, 20] and Rashba effect at interfaces[18] can be used to achieve larger spin-to-charge conversion efficiencies with high resistivity and further improve the voltage readout of the miniaturized device as well as the current output to allow cascading to the next one, two ingredients which are essential for logic operations in computational applications[4,15].

**Acknowledgments**

The authors acknowledge Roger Llopis and Ralph Gay for technical assistance with the sample fabrication, and thank Dr. Shehrin Sayed for fruitful discussions on the transmission line model and the equivalent circuit. V.T.P. thanks Dr. Laurent Vila for fruitful discussions on the local spin detection/injection technique. This work is supported by Semiconductor Research Corporation under Project No. 2017-IN-2744 and by the Spanish MINECO under the Maria de Maeztu Units of Excellence Programme (MDM-2016-0618) and under Projects No. MAT2015-65159-R, RTI2018-094861-B-100, and MAT2017-82071-ERC. V.T.P. acknowledges postdoctoral fellowship support "Juan de la Cierva-incorporación" program by the Spanish MINECO (Grant No. FJCI-2017-34494). E.S. thanks the Spanish MECD for a Ph.D. fellowship (Grant No. FPU14/03102).

**Author contributions**

V.T.P. and F.C. conceived the study. V.T.P. and I.G. performed the experiments. V.T.P., I.G., S.M., W.Y.C., D.E.N., E.S., C-C.L., T.G., I.Y., S.M., L.E.H. and F.C. analyzed the data and discussed the experiments. V.T.P. derived the equations from the 1D spin diffusion model. V.T.P. and A.M. performed the 3D-FEM simulation based on the spin diffusion model. V.T.P., S.M. and F.C. wrote the manuscript. All the authors contributed to the scientific discussion and manuscript revision. F.C. supervised the work.

**Competing interests**

Authors declare no competing interests.

**Additional Information**

Extended Data and Supplementary Information is available for this paper.

# Methods

**Nanofabrication**

The devices were fabricated on SiO$_2$/Si substrates with multiple-step e-beam lithography, metal deposition, and lift-off process. In most cases, the spin Hall nanowire (Pt) is patterned before the ferromagnetic electrode (CoFe). The Pt layer is deposited by magnetron sputtering using conditions (1.3 Å/s, 80 W of power, $1.0\times10^{-8}$ mTorr of base pressure, 3 mTorr of Ar pressure) that favor high resistivity[26], whereas $Co_{50}Fe_{50}$ is e-beam evaporated (at 0.4 Å/s and $5.0\times10^{-7}$ mTorr). The CoFe electrode is patterned to have a single magnetic domain with the easy axis of the magnetization along the wire (*x*-direction). Between the two depositions, an Ar-ion milling process (out-of-plane Ar-ion flow of 15 sccm, an acceleration voltage of 50 V, a beam current of 50 mA, and a beam voltage of 300 V) is performed for ~30 s to ensure a transparent CoFe/Pt interface.

**Measurements**

Electronic transport measurements are performed in a Physical Property Measurement System from Quantum Design, using a 'DC reversal' technique with a Keithley 2182 nanovoltmeter and a 6221 current source at temperatures ranging from 10 to 300 K. The applied current $I_c^{app}$ applied for the measurements is 10, 20, or 50 µA, depending on the size of the device. We apply in-plane and out-of-plane magnetic fields with a superconducting solenoid magnet by rotating the sample using a rotatable sample stage.

**Data availability**

The data that supports the findings of this study are available from the corresponding author on reasonable request.

**Extended Data Table 1 | Summary of devices and device parameters**

| Sample | $\rho_{Pt}$ ($\mu\Omega cm$) | $\rho_{CoFe}$ ($\mu\Omega cm$) | $t_{Pt}$ (nm) | $t_{CoFe}$ (nm) | $w_{Pt}$ (nm) | $w_{CoFe}$ (nm) | $2\Delta_{(I)SHE}$ ($m\Omega$) |
|---|---|---|---|---|---|---|---|
| LD1 | 174.0 | 320.0 | 6 | 10 | 65 | 50 | 300 ± 50 |
| LD2 | 96.0 | 42.0 | 10 | 19 | 160 | 210 | 14.8 ± 0.5 |
| LD3a | 95.7 | 116.8 | 10 | 19 | 270 | 70 | 13.8 ± 1.2 |
| LD3b | 47.7 | 108.9 | 10 | 19 | 270 | 560 | 9.0 ± 0.6 |
| LD3c | 93.7 | 110.0 | 10 | 19 | 270 | 550 | 13.3 ± 2.0 |
| LD3d | 95.9 | 112.9 | 10 | 19 | 270 | 270 | 12.6 ± 1.0 |
| LD3e | 97.6 | 100.2 | 10 | 19 | 270 | 110 | 11.6 ± 2.2 |
| LD3f | 106.1 | 116.9 | 10 | 19 | 270 | 220 | 12.1 ± 1.4 |
| LD3g | 106.1 | 116.9 | 10 | 19 | 270 | 220 | 10.8 ± 1.4 |
| LD4a | 71.2 | 104.9 | 10 | 19 | 95 | 240 | 30.6 ± 3.1 |
| LD4b | 73.0 | 116.0 | 10 | 19 | 135 | 220 | 23.7 ± 2.7 |
| LD4c | 74.9 | 114.0 | 10 | 19 | 230 | 240 | 14.4 ± 1.2 |
| LD4d | 70.1 | 84.6 | 10 | 19 | 335 | 240 | 6.0 ± 1.3 |
| LD4e | 72.7 | 113.8 | 10 | 19 | 450 | 230 | 5.8 ± 0.6 |
| LD5 | 56.6 | 91.3 | 10 | 19 | 110 | 300 | 32.0 ± 0.6 |
| LD6[*] | 90.0 | 830.0 | 20 | 14 | 130 | 60 | 35 ± 5 |
| LD7[*] | 30.3 | 142.6 | 20 | 16 | 250 | 90 | 5.0 ± 0.2 |
| LD8 | 225 | 92.5 | 8 | 15 | 35 | 55 | 140 ± 20 |

Resistivities, dimensions, and spin Hall signals at 300 K for different devices. The errors of the spin Hall signals are estimated from the measurement noise. The devices are fabricated on different $SiO_2$/Si substrates (LD1–LD8). [*]These samples are fabricated with opposite stacking order, *i.e.,* with Pt on top of CoFe.

# Supplementary Information

**Supplementary Note 1: Reproducibility of the giant spin Hall signal and signal offset**

A giant spin Hall signal with a hysteresis loop following the magnetization state was obtained in another device fabricated on the same substrate as device LD1 presented in the main text. Supplementary Fig. 1a shows a SEM image of the nanostructure and the measurement configuration for ISHE. The dimensions of the Pt T-shaped nanostructure are identical to the ones in device LD1 ($t_{Pt}$ = 6 nm, $t_{CoFe}$ = 10 nm and $w_{Pt}$ = 60 nm), whereas the CoFe electrode is wider, $w_{CoFe}$ = 300 nm (cf. $w_{CoFe}$ = 50 nm in LD1). The transverse resistance $R_{ISHE} = V_{ISHE}/I_C^{app}$ as a function of the magnetic field at 10 K is presented in Supplementary Fig. 1b. As expected from Supplementary Equation 4, the spin Hall signal in this device is the same as for device LD1 shown in Figs. 1d and 1e of the main text.

Contrary to magnetoresistance-based effects, the signal offset of the transverse resistance $R_{ISHE}$ must be zero in a symmetric nanostructure. Indeed, it is very low in this device (~ −0.1 Ω). The non-zero offset in device LD1 (~67 Ω, see Figures 1d and 1e of the main text) can be explained by the misalignment (~20 nm in *y*-direction) between the two steps of the nanofabrication during the e-beam lithography (see SEM images in Figures 1b and 1c).

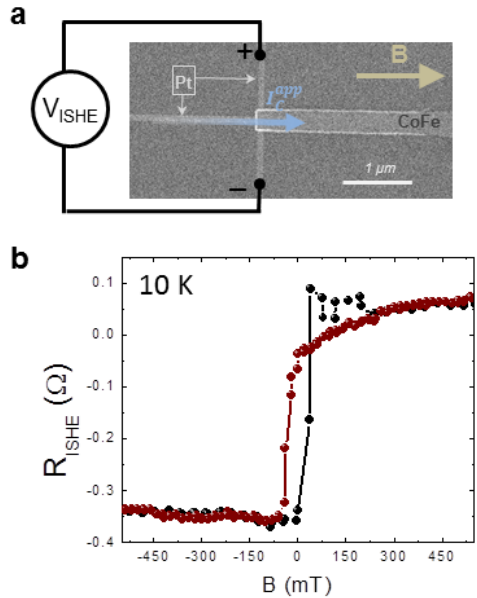

**Supplementary Figure 1: a**, SEM image of the second spin-to-charge conversion device in substrate LD1 and the measurement configuration for ISHE. **b**, Transverse resistance $R_{ISHE}$ as a function of the magnetic field (trace and retrace) measured at 10 K.

**Supplementary Note 2: 1D analytical spin diffusion model for the spin-to-charge conversion device used in this work**

The expressions of the output of the ISHE setup (Figure. 1a in the main text) in terms of voltage and in terms of charge current are based on the two-current diffusion model[40]. Starting with the assumption that the electrochemical potentials for spin-up and spin-down are continuous through the transparent interface of CoFe/Pt, the expression of the charge current output produced by the ISHE ($I_C^{ISHE}$) in the $y$-direction can be taken from Ref. 25 (Equation S7 of its Supplementary information). By adapting the expression to our CoFe/Pt nanostructure, we obtain:

$$I_C^{ISHE} = P_{CoFe}\theta_{SH}\lambda_{Pt} \frac{1 - \frac{1}{\cosh(t_{Pt}/\lambda_{Pt})}}{\tanh(t_{Pt}/\lambda_{Pt}) + \frac{\lambda_{Pt}\rho_{Pt}}{\lambda_{CoFe}\rho_{CoFe}^*}\tanh(t_{CoFe}/\lambda_{CoFe})} \int j_{cz} dx, \quad (1)$$

with $\rho_{CoFe}^* = \rho_{CoFe}/(1 - P_{CoFe}^2)$, where $\theta_{SH}$, $P_{CoFe}$, $\lambda_{Pt,CoFe}$, $t_{Pt,CoFe}$ and $\rho_{Pt,CoFe}$ are the spin Hall angle of Pt, the spin polarization of CoFe, the spin diffusion length, the thickness and the resistivity, respectively. The subscripts denote the materials. $j_{cz}$ is the applied charge current density in the $z$-direction. We assume the current will contribute homogeneously in the $y$-direction, such that $I_C^{app} = \iint j_{cz} dx dy = w_{CoFe} \int j_{cz} dx$ (see Supplementary Fig. 2 for the definition of the dimensions). Hence, the integral in Supplementary Equation 1 can be written as $\int j_{cz} dx = I_C^{app}/w_{CoFe}$ and the expression for the charge current output in the transverse Pt wire becomes:

$$I_C^{ISHE} = P_{CoFe}\theta_{SH}\lambda_{Pt} \frac{1 - \frac{1}{\cosh(t_{Pt}/\lambda_{Pt})}}{\tanh(t_{Pt}/\lambda_{Pt}) + \frac{\lambda_{Pt}\rho_{Pt}}{\lambda_{CoFe}\rho_{CoFe}^*}\tanh(t_{CoFe}/\lambda_{CoFe})} \frac{1}{w_{CoFe}} I_C^{app}. \quad (2)$$

The open circuit condition for the transverse Pt wire allows us to write the voltage output as $\Delta V_{ISHE} = R_T I_C^{ISHE}$, where $R_T$ is the transverse resistance:

$$R_T = \frac{w_{CoFe}}{\left(\frac{t_{CoFe}}{\rho_{CoFe}} + \frac{t_{Pt}}{\rho_{Pt}}\right)w_{Pt}}, \quad (3)$$

so that the voltage output normalized to the applied current (*i.e.*, the spin Hall signal) is then given by:

$$\frac{\Delta V_{ISHE}}{I_C^{app}} = \frac{P_{CoFe}\theta_{SH}\lambda_{Pt}}{\left(\frac{t_{CoFe}}{\rho_{CoFe}} + \frac{t_{Pt}}{\rho_{Pt}}\right)w_{Pt}} \times \frac{1 - \frac{1}{\cosh(t_{Pt}/\lambda_{Pt})}}{\tanh(t_{Pt}/\lambda_{Pt}) + \frac{\lambda_{Pt}\rho_{Pt}}{\lambda_{CoFe}\rho_{CoFe}^*}\tanh(t_{CoFe}/\lambda_{CoFe})}. \quad (4)$$

The spin Hall signal is maximum for the case of $t_{Pt}/\lambda_{Pt} = 2$. For systems composed of materials with much shorter spin diffusion lengths than their thicknesses, which is the case in this work ($t_{Pt}/\lambda_{Pt} \gg 1$, $t_{CoFe}/\lambda_{CoFe} \gg 1$), the spin Hall signal can be simplified to:

$$\frac{\Delta V_{ISHE}}{I_C^{app}} = \frac{1}{\left(\frac{t_{CoFe}}{\rho_{CoFe}} + \frac{t_{Pt}}{\rho_{Pt}}\right)w_{Pt}} \times \frac{P_{CoFe}\theta_{SH}\lambda_{Pt}}{1 + \frac{\lambda_{Pt}\rho_{Pt}}{\lambda_{CoFe}\rho_{CoFe}^*}}, \quad (5)$$

and the normalized charge current output can be simplified to:

$$\frac{I_C^{ISHE}}{I_C^{app}} = \frac{1}{w_{CoFe}} \times \frac{P_{CoFe}\theta_{SH}\lambda_{Pt}}{1+\frac{\lambda_{Pt}\rho_{Pt}}{\lambda_{CoFe}\rho_{CoFe}^*}}. \qquad (6)$$

To understand the spin-to-charge current conversion in our devices given by Supplementary Equations 5 and 6, we have defined in the main text the "geometrical" factor $G = \frac{1}{\left(\frac{t_{CoFe}}{\rho_{CoFe}}+\frac{t_{Pt}}{\rho_{Pt}}\right)w_{Pt}}$ and the "efficiency" factor $\lambda_{eff} = \frac{P_{CoFe}\theta_{SH}\lambda_{Pt}}{1+\frac{\lambda_{Pt}\rho_{Pt}}{\lambda_{CoFe}\rho_{CoFe}^*}}$, leading to Equations 2 and 3 of the main text.

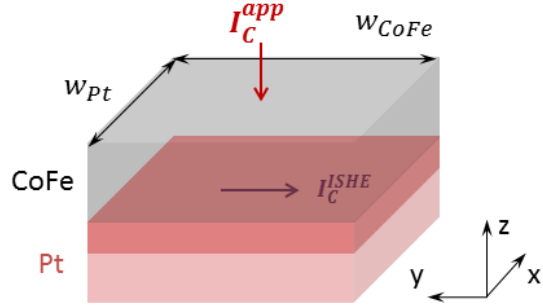

**Supplementary Figure 2:** The 1D analytical model is described by a CoFe/Pt bilayer slab corresponding to the intersection region of the real spin-to-charge conversion device, where the spin-to-charge current conversion occurs.

**Supplementary Note 3: 3D numerical spin diffusion model for the spin-to-charge conversion device used in this work**

Finite element method (FEM) simulations are also performed within the framework of the two-current drift-diffusion model, with the collinear magnetization of the FM electrode along the easy axis. The geometry construction and 3D-mesh were elaborated using the free software GMSH with the associated solver GETDP[41] for calculations, post-processing and data flow control. The geometry and the mesh of the simulation are shown in Supplementary Fig. 3a. Since the top surface of Pt is cleaned by out-of-plane Ar-ion flow (see section in Materials and Methods for the nanofabrication in the main text), only the top FM/NM interface is assumed to be transparent, without spin memory loss, and thus the electrochemical potentials for spin-up and spin-down are continuous through the entire device. The lateral FM/NM interface is assumed to be insulating because the side surface of Pt is not etched. The Neumann boundary conditions are applied. The model is properly designed by taking into account that: i) the nanowires are long enough (*i.e.,* longer than 3 times the spin diffusion lengths in the materials, and much longer than the nanowire widths) such that the spin current vanishes (in the ISHE configuration) at their ends; ii) the mesh size in the vicinity of the interface (where the spin-to-charge current conversion takes place) is set smaller in order to ensure that the SOC-based effects are calculated properly. Further details concerning the FEM model can be found in Ref. 25.

We simulate devices LD1 and LD2 using this model. The results, which also include the AHE contribution (Supplementary Note 4), are plotted in Figures 2c and 2d of the main text.

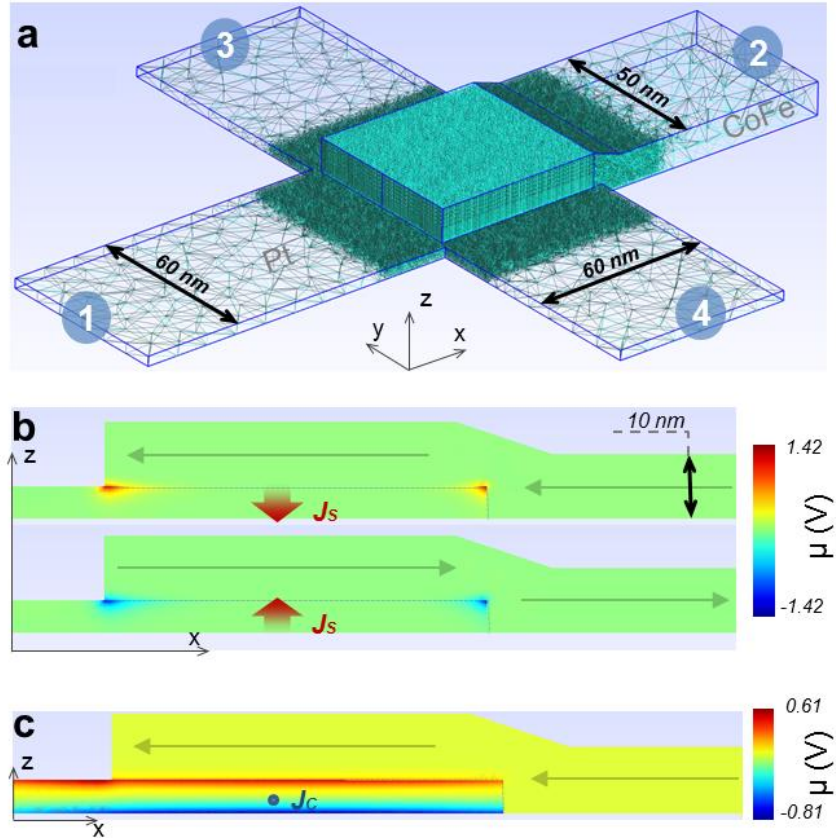

**Supplementary Figure 3: a**, Geometry and mesh of the model used for retrieving the spin Hall signals. The numbers indicate the terminals for the current injection and voltage detection. Panels b and c present the $x - z$ cross-section contour of the spin accumulation profile (i.e., half the difference between the electrochemical potentials of the majority and minority spins) for the ISHE and SHE configurations, respectively. The spin current is proportional to the gradient of the spin accumulation. **b**, In the ISHE measurement configuration, the charge current flows from terminal 1 to 2 so that a spin-polarized current is injected along the $z-$direction though the CoFe/Pt interface. The polarization of the injected spin current depends on the magnetization state of the ferromagnetic electrode (along $+x$ or $-x$). In the case of the magnetization oriented along $+x$ (top panel), the spin accumulation is positive so that the $z-$component of the spin current is along $-z$, which leads to the appearance of a positive ISHE signal. In the case of the magnetization oriented along $-x$ (bottom panel), the spin accumulation is negative so that the $z-$component of the spin current is along $+z$, which leads to the appearance of a negative ISHE signal. **c**, In the SHE measurement configuration, the charge current flows from terminal 4 to 3, inducing a pure spin current along the $z-$direction. The resulting spin accumulation at the top surface of the Pt is then probed by the ferromagnetic electrode, leading to a voltage difference between terminal 1 and 2. Since all Equations used for the simulations are linear, the injected current has been set to the unit current, *i.e.*, 1 A.

# Supplementary Note 4: Estimation of the anomalous Hall effect contribution to the spin Hall signal

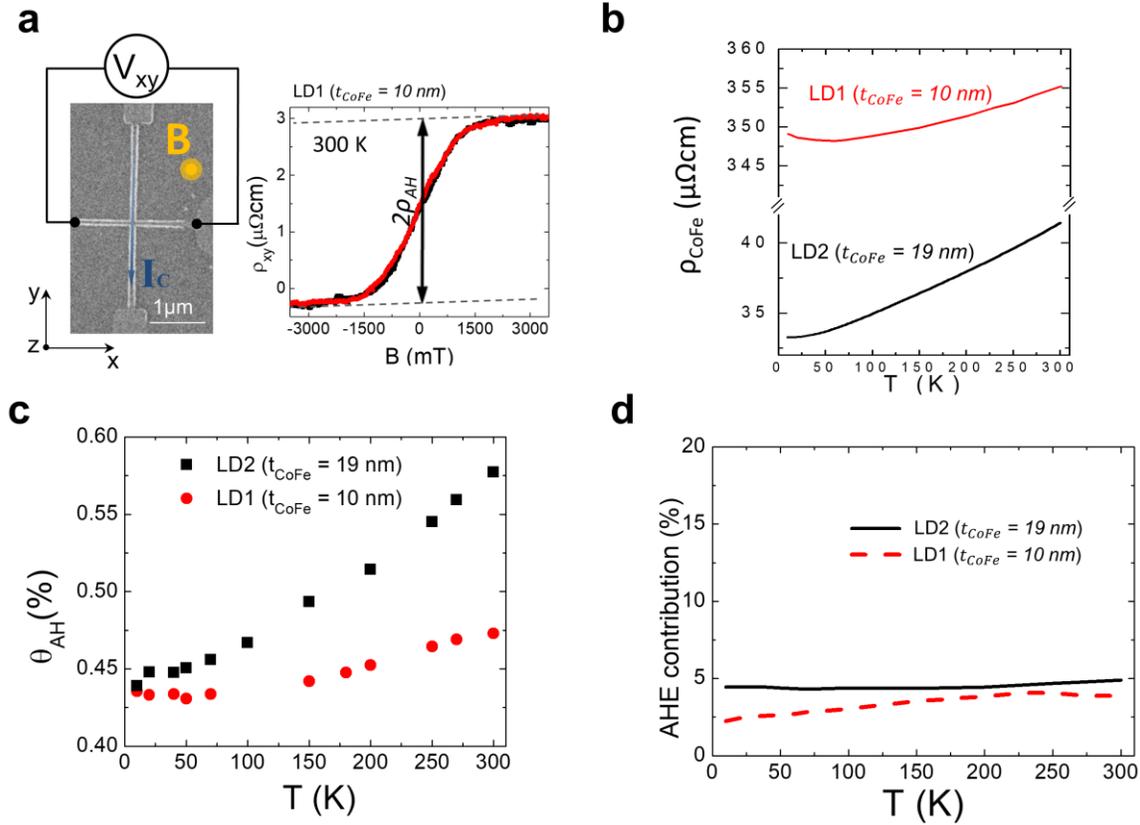

**Supplementary Figure 4: a**, Setup of the AHE measurement pictured on a CoFe cross of device LD1 (left) and corresponding transverse resistivity as a function of the out-of-plane magnetic field measured at room temperature (right). The black arrow indicates the anomalous Hall resistivity, $\rho_{AH}$. **b**, Resistivity of CoFe in devices LD1 ($t_{CoFe}$ = 10 nm) and LD2 ($t_{CoFe}$ = 19 nm) as a function of temperature. **c**, Anomalous Hall angle of devices LD1 and LD2 as a function of temperature obtained from the measurements shown in **a** and **b**. **d**, AHE contribution to the spin Hall signal as a function of temperature for devices LD1 and LD2, which is calculated by the 3D-FEM model (Supplementary Note 3).

In the ISHE measurement configuration (Figure 1a of the main text), the applied charge current flows along the $z$−direction at the intersection region. With the in-plane magnetization of the CoFe electrode being along the $x$−direction, a voltage signal originating from the anomalous Hall effect (AHE) of CoFe could appear along the $y$−direction. Likewise, in the reciprocal SHE measurement (Figure 1c of the main text), a part of the applied current in Pt along the $y$−direction would flow through the CoFe (with magnetization along $x$), giving rise to a vertical AHE electric field. This voltage gradient along the $z$−direction would be probed differently by the Pt terminal and the CoFe terminal, leading to the AHE signal. In both cases, the AHE signal will appear simultaneously with the spin Hall signal and with the same shape of the hysteresis loop. In general, this contribution is very small and can be neglected in most cases due to the geometrical configuration. A minimal AHE contribution can be achieved by choosing the optimal thickness ratio between the CoFe and Pt layers, $t_{FM} \approx 2t_{NM}$.[42] In this section, we quantify the contribution of the AHE to the measured spin Hall signal for devices LD1 and LD2.

The AHE of CoFe is measured using a Hall configuration at a cross fabricated in the same devices as where the spin Hall signals are measured, as shown in the left panel of Supplementary Fig. 4b. The transverse resistivity $\rho_{xy} = (V_{xy}/I_c)t_{CoFe}$ as a function of the out-of-plane magnetic field is shown in the right panel of Supplementary Fig. 4b. The anomalous Hall resistivity, $\rho_{AH}$, is calculated from the high field extrapolations to zero field, as indicated in the plot. The anomalous Hall angle, $\theta_{AH} = \rho_{AH}/\rho_{CoFe}$, as a function of temperature for devices LD1 and LD2 is plotted in Supplementary Fig. 4c, where the resistivity of CoFe has been independently measured for each device (see Supplementary Fig. 4a). The obtained $\theta_{AH}$ is used as an input for the 3D-FEM simulations presented in Supplementary Note 3. From the 3D simulations, we can extract the contribution of the AHE. The ratio between the AHE signal calculated by the 3D-FEM simulation and the experimental spin Hall signal is plotted in Supplementary Fig. 4d.

**Supplementary Note 5: Planar Hall effect**

In this note, we estimate the contribution of the planar Hall effect (PHE) of the CoFe electrode to the spin Hall signal in a Pt/CoFe sample. The PHE can contaminate the spin Hall signal only if the magnetization of the CoFe is not parallel to the applied current direction. We performed angular dependent measurements with the electrical probes for the ISHE configuration in device LD7b ($t_{Pt} = 20$ nm, $t_{CoFe} = 16$ nm and $w_{Pt} = 260$ nm) as shown in Supplementary Fig. 5a. The ISHE signal is proportional to the cosine of the azimuthal angle $\alpha$ (the angle between the applied current density and the CoFe magnetization), whereas the PHE is proportional to sin($2\alpha$). Thus, in principle, the PHE does not play any role in the ISHE signal because the experiments performed in the main text correspond to a measurement configuration of $\alpha = 180º$ and $\alpha = 0º$. The PHE can appear only by a misalignment $\alpha_0$ between the applied magnetic field and the easy axis of the CoFe electrode in the measurement setup. For the angular dependent measurement, the transverse resistance can be described by:

$$R_T = \frac{V_{ISHE} + V_{PHE}}{I_c^{app}} = a_{ISHE}\cos(\alpha + \alpha_0) + a_{PHE}\sin(2\alpha + 2\alpha_0) + R_{baseline} \qquad (7)$$

where $a_{ISHE}$ and $a_{PHE}$ are constants indicating the amplitude of ISHE and PHE, respectively. $\alpha_0$ is the angle correction.

The sample is rotated in the $x - y$ plane (see Supplementary Fig. 5a) in a magnetic field **B** higher than the saturation field of the CoFe electrode to ensure that the magnetization of the CoFe electrode is directed along **B**. The results are shown in Supplementary Fig. 5b. We have fitted Supplementary Equation 7 to the experimental data and extracted the individual signals of the PHE and the ISHE. The fitting parameters are shown in Supplementary Table 1. As expected, the spin Hall signal is maximized while the PHE signal is minimized at 180º and 360º.

Therefore, the spin Hall signal extracted from the angular dependence coincides with the amplitude of the signal measured in the configuration used in the main text (see Supplementary Fig. 5c). In other words, the parameter $a_{ISHE} = 1.2$ mΩ is in a good agreement with the amplitude of the spin Hall signal shown in Supplementary Fig. 5c. It is clear that the PHE can contaminate the spin Hall signal only by an imperfect alignment when mounting the sample for the ISHE measurement. The misalignment is given by $\alpha_0$ in Supplementary Equation 7. From our fit, we obtain a misalignment angle of $\alpha_0 = 2.4°$, so that the planar Hall contribution is $\Delta R_{PHE} = a_{PHE} \sin(2\alpha_0) = -0.06$ mΩ. This quantity is very small, which allows us to conclude that the contribution of the PHE to the spin Hall signal (~5%) is negligible.

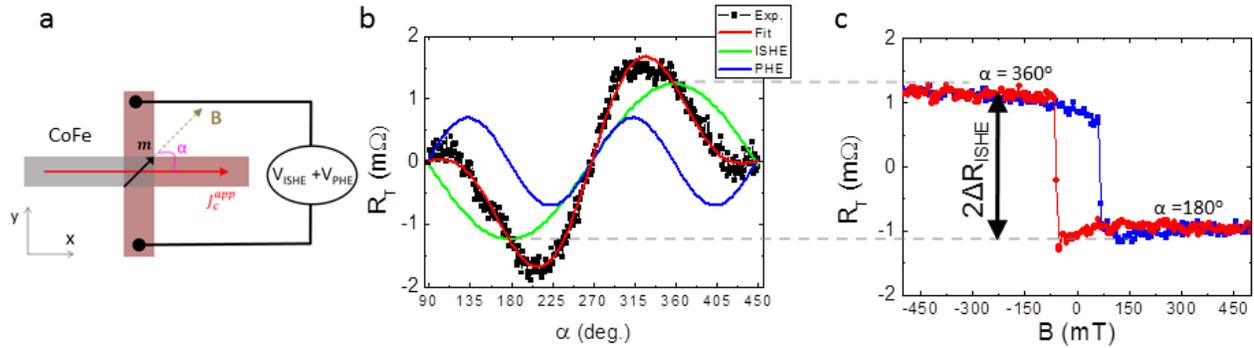

**Supplementary Figure 5**. **a**, Sketch of the angular dependent measurement of the spin Hall signal performed in device LD7b ($t_{Pt} = 20$ nm, $t_{CoFe} = 16$ nm and $w_{Pt} = 260$ nm). **b**, Transverse resistance $R_T$ as a function of the azimuthal angle ($\alpha$) between the applied current density and the magnetization measured at 10 K. An external magnetic field ($B = 3.2$ T) is applied to orient the magnetization. The experimental data is shown as black solid squares and the fitting curve using Supplementary Equation 7 is shown as a red line, which can be decomposed into the ISHE (green line) and the PHE (blue line) contributions. **c**, $R_T$ as a function of the magnetic field applied along the easy axis of the CoFe electrode at 10 K. The two magnetization orientations corresponding to $\alpha = 180°$ and $\alpha = 360°$ are indicated. Blue squares correspond to trace of the magnetic field and red circles to retrace. The stacking order of this device is opposite compared to the ones in the main text (*i.e.*, Pt/CoFe versus CoFe/Pt stacks) and, therefore, the sign of the spin Hall signal is reversed. The baseline of the signals is removed in **b** and **c** after fitting.

**Supplementary Table 1**. Fitting parameters of the angular dependent transverse resistance measurement in device LD7b at 10 K.

| $a_{PHE}$ | $a_{ISHE}$ | $\alpha_0$ | $R_{baseline}$ |
|---|---|---|---|
| -0.7 mΩ | 1.2 mΩ | 2.4° | 532.48 mΩ |

## Supplementary Note 6: Thermal effects

Spin caloritronic effects (longitudinal spin Seebeck effect, planar Nernst effect) refer to the generation of spin voltage as a result of a temperature gradient. In our devices, a temperature gradient is possible due to the applied current through the CoFe/Pt interface and the generated spin voltage could be picked up when we measure with the ISHE configuration. To rule out the possibility of thermal contributions in our spin signals, we checked the dependence of the spin Hall signal with the applied current in device LD7b. The experiments show clearly a linear relationship between the ISHE voltage and the applied current (Supplementary Fig. 6), whereas the thermal effects caused by a temperature gradient are proportional to the square of the applied current. Therefore, we confirm that no thermal effects contaminate the obtained spin Hall signals.

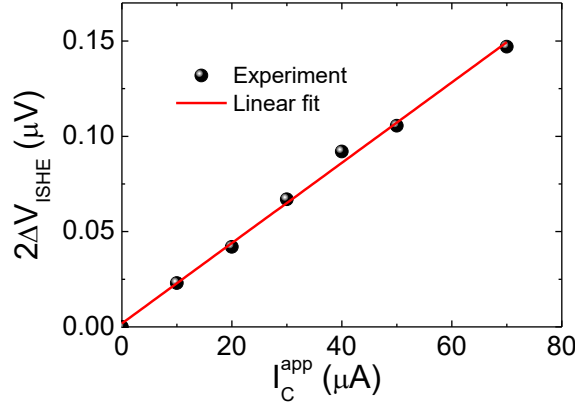

**Supplementary Figure 6.** Spin Hall voltage as a function of the applied current measured at 10 K in device LD7b ($t_{Pt}$ = 20 nm, $t_{CoFe}$ = 16 nm and $w_{Pt}$ = 260 nm). Red line is a linear fit to the experimental data. The applied current in the experiments shown in the main text vary from 10 to 40 µA.

**Supplementary Note 7: Negative spin Hall signal in a CoFe/Ta device**

In order to further confirm that the main origin of our signal is the spin Hall effect, we performed a control experiment using Ta, a spin Hall metal with a negative spin Hall angle Ta. Supplementary Fig. 7a shows a scanning electron microscopy (SEM) image of the CoFe/Ta device with the measurement configuration for the ISHE. A T-shaped nanostructure made of Ta replaces the one made of Pt in Fig. 1b of the main text. The widths of Ta and CoFe wires are $w_{Ta}$ = 90 nm and $w_{CoFe}$ =160 nm, respectively. The thicknesses of Ta and CoFe wires are $t_{Ta}$ = 10 nm and $t_{CoFe}$ =15 nm, respectively. The resistivity of Ta is $\rho_{Ta}$=1600 µΩcm. In order to inject the spin current from CoFe to this high resistivity Ta, a thin layer of AlO$_x$ is deposited at the interface. A high value of the product between the interface area ($A_i$) and the interface resistance ($R_i$), $R_i A_i$ = 11.5 Ωµm$^2$, is measured in a cross fabricated in the same device. The transverse resistance $R_{ISHE}$ as a function of the magnetic field, measured at 10 K, is shown in Supplementary Fig. 7b. It shows a spin Hall signal with reversed sign compared to the ones in Pt presented in the main text, as expected from the negative spin Hall angle of Ta. The signal, $2\Delta R_{ISHE} = -3.4 \pm 0.3$ Ω, is even higher than the ones reported in the Pt-based devices, mainly because of the high resistivity of Ta leading to a high geometrical factor, which we calculate to be $G^*$ = 17.8 Ω/nm. We can then extract $\lambda^*_{eff}$ = 0.10 nm for the Ta/CoFe system, which is two times larger than for the Pt/CoFe devices. Note that $G^*$ and $\lambda^*_{eff}$ parameters are defined for the model with a high interface resistance (adapted from Ref. 27 in the main text), with $G^* = \rho_{Ta} t_{Ta}/w_{Ta}$ and $\lambda^*_{eff} = P_{CoFe}\lambda_{Ta}\theta_{Ta}$. Although they are defined slightly different from $G$ and $\lambda_{eff}$ defined in the main text for the case of a transparent interface, they have the equivalent role.

This control experiment confirms that all the results presented in this work are primarily due to the spin Hall effect. Furthermore, it shows that higher spin Hall signals can be obtained by adapting the local spin injection technique introduced here to other spin-orbit materials that increase the internal resistance of the ISHE current source ($G$) and/or the efficiency factor ($\lambda_{eff}$), as discussed in the last paragraph of the main text before the conclusions. Our high record amplitudes of the experimental spin Hall signals in such a simple metallic nanostructure will sure inspire and invigorate many research groups to develop and use new materials with larger $\lambda_{eff}$ (including more exotic materials such as topological insulators, Weyl semimetals, or 2D materials) towards spintronics devices for computational applications, and particularly for our recent proposal of the scalable energy-efficient MESO logic (Ref. 4 in the main text).

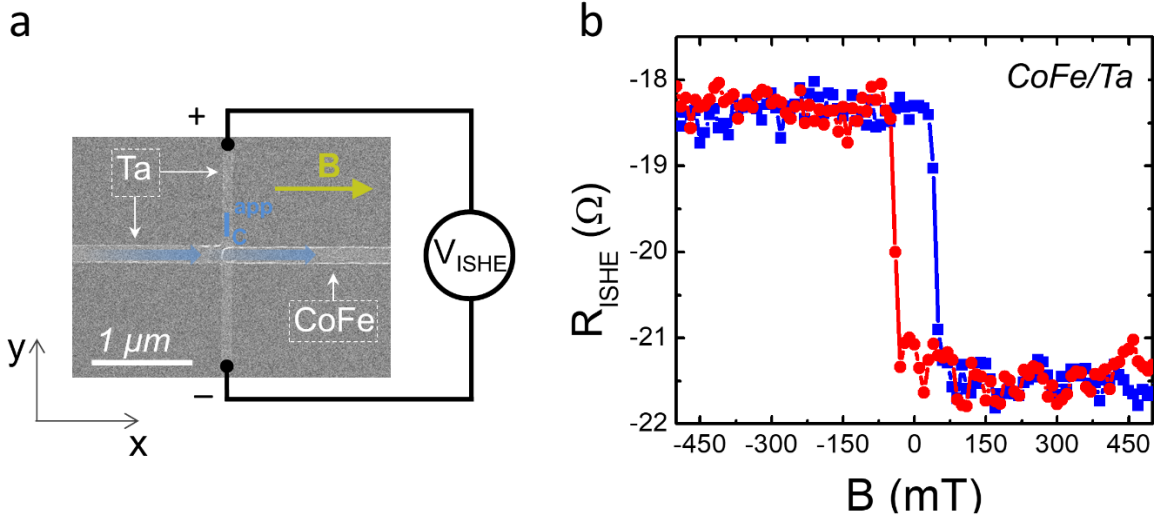

**Supplementary Figure 7**. **a**, SEM image of a CoFe/Ta device with the measurement configurations for the ISHE. **b**, Transverse resistance ($R_{ISHE}$) as a function of the magnetic field (trace in blue and retrace in red) measured in the CoFe/Ta device at 10 K. The reverse sign of the spin Hall signal as compared to the signal of CoFe/Pt devices is observed.

**Supplementary Note 8: Width dependence of the spin Hall signal**

To study the width dependence of the spin Hall signal in the local device used in this work (see Fig. 1a of the main text for a sketch), we fabricated two sets of devices on two different SiO$_2$/Si substrates (LD3 and LD4). LD3 contains the set of devices with varying $w_{Pt}$ and LD4 the set of devices with varying $w_{CoFe}$ (see Extended Data Table 1 for the complete set of device parameters).

The experimental spin Hall signals as a function of the width of the CoFe and the Pt wire are plotted in Supplementary Fig. 8a and 8b, respectively. We observed that the spin Hall signal is independent of $w_{CoFe}$, whereas it is inversely proportional to $w_{Pt}$, as predicted by Supplementary Equation 5 (Equation 2 of the main text). In this regard, the fitting parameter $a = 3.2 \pm 0.5$ Ωnm obtained from the plot in Supplementary Fig. 8b should correspond to $a = \frac{1}{\left(\frac{t_{CoFe}}{\rho_{CoFe}} + \frac{t_{Pt}}{\rho_{Pt}}\right)} \times \frac{P_{CoFe}\theta_{SH}\lambda_{Pt}}{1+\frac{\lambda_{Pt}\rho_{Pt}}{\lambda_{CoFe}\rho^*_{CoFe}}}$ for this set of devices (LD4). Indeed, if we calculate $a$ by taking into account the known parameters of the set of devices LD4 (from Extended Data Table 1) and the material parameters (from Refs. 25, 29 and 30), $a$ ranges from 3.21 to 3.85 Ωnm (arising from the dispersion of the resistivities in the set of devices).

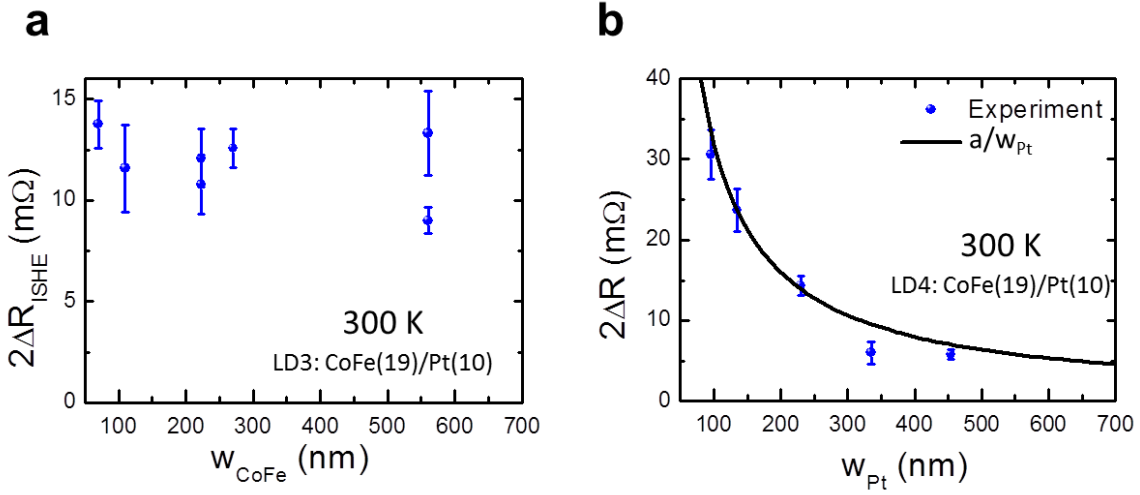

**Supplementary Figure 8: a**, Spin Hall signal as a function of the width of the CoFe electrode measured at room temperature in the set of devices LD3, where the width of the Pt wire is kept as $w_{Pt} = 270$ nm. **b**, Spin Hall signal as a function of the width of the Pt wire measured at room temperature in the set of devices LD4, where the width of CoFe wire is kept as $w_{CoFe} = 240$ nm. The black solid line is a fit of $a/w_{Pt}$ to the experimental data. All devices have thicknesses of $t_{Pt} = 10$ nm and $t_{CoFe} = 19$ nm.

**Supplementary Note 9: Equivalent circuit of the inverse spin Hall effect configuration**

Based on the transmission model for materials with spin-momentum locking (including bulk spin Hall effect) shown in Ref. 36, the equivalent circuit of the ISHE configuration under open circuit condition can be described as a current source ($I_C^{ISHE}$) with an internal resistance ($R_T$) (Supplementary Fig. 9a):

i) The current source $I_C^{ISHE}$ (see Supplementary Equation 6 and Equation 3 of the main text) is proportional to the "efficiency" factor $\lambda_{eff}$ and inversely proportional to the CoFe wire width ($w_{CoFe}$). This can be understood from the nature of the ISHE, where the spin Hall angle is a constant value given by the rate of current densities, i.e., $\theta_{SH} = [\hbar/e] J_C^{ISHE}/J_S$. Because of the transverse geometry of the SHE effect, the areas defining the two current densities are different. Whereas $w_{CoFe} \times w_{Pt}$ defines the area of the injected spin current density ($J_S$), $\lambda_{Pt} \times w_{Pt}$ defines the area of the output charge current density ($J_C^{ISHE}$), see upper sketch in Supplementary Fig. 9b. The spin-polarized electrons injected into the Pt will deflect at each scattering event a constant angle $\alpha$ [$\tan(\alpha) = \theta_{SH}$]. As far as the spins are conserved (i.e., for a depth $\lambda_{Pt}$), the deflection will be towards the same side, acquiring a transverse velocity that leads to the ISHE current. Therefore, $\lambda_{eff}$ [which accounts for the spin-to-charge conversion efficiency in Pt (the $\theta_{SH}\lambda_{Pt}$ product), the spin polarization of CoFe ($P_{CoFe}$), and the spin resistance mismatch of the two materials ($\lambda_{Pt}\rho_{Pt}/\lambda_{CoFe}\rho_{CoFe}^*$)] can be seen as the effective length in the $y$-direction along which the ISHE current is generated (see lower sketch in Supplementary Fig. 9b). Ideally, for a system with a $\lambda_{eff}$ large enough to match the ferromagnet width, the rate of the transverse charge current output ($I_C^{ISHE}$) to the applied charge current ($I_C^{app}$) could reach one (see Equation 3 of the main text).

ii) The internal resistance is given by the transverse resistance $R_T$ of the Pt/CoFe intersection region (see Supplementary Equation 3). Here, $w_{CoFe}$ plays the role of the length of the internal resistance and the "geometrical" factor $G$ is the transverse resistance per unit length (Supplementary Fig. 9c).

Taking into account these two elements, the output voltage is simply given by $R_T I_C^{ISHE}$ (see Supplementary Equation 5), which is inversely proportional to the Pt wire width $w_{Pt}$, but then becomes independent of $w_{CoFe}$.

Since each element (current source and internal resistance) has a different scaling law, the ISHE outputs in current and in voltage can be independently optimized by tuning the different lateral dimensions, as we demonstrate in Figure 3 of the main text.

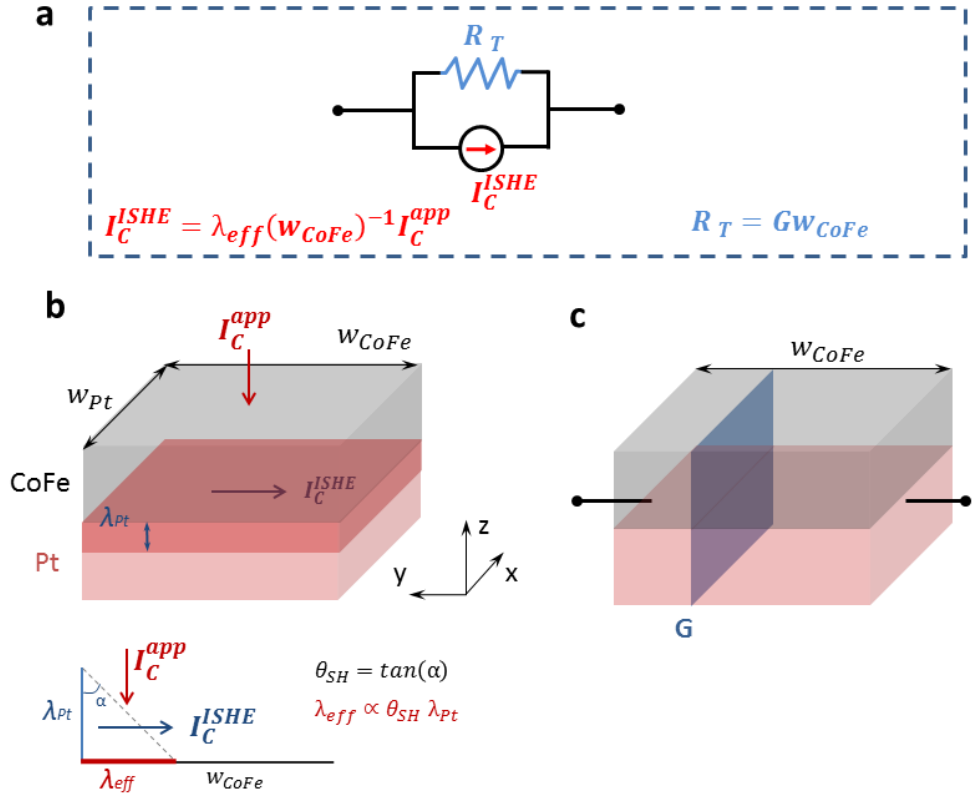

**Supplementary Figure 9: a**, Equivalent circuit of the ISHE measurement under open circuit condition: the ISHE can be described as a current source with an internal resistance. The measured voltage is the product of the two elements. **b**, Upper panel: Sketch of the active part of the device acting as the current source, where the relevant dimensions are tagged. Lower panel: Geometrical interpretation of the effective length, $\lambda_{eff}$. **c**, Sketch of the internal resistance, effectively a homogeneous slab with length $w_{CoFe}$ and a transverse resistance per unit length, $G$.